# Dielectric properties of Poly(vinylidene fluoride)/ $CaCu_3Ti_4O_{12}$ nanocrystal composite thick films.

P.Thomas[a,b], S. Satapathy[c], K.Dwarakanath,[a] and K.B.R.Varma.[b]∗

[a] Dielectric Materials Division, Central Power Research Institute, Bangalore:560080, India

[b] Materials Research Centre, Indian Institute of Science, Bangalore: 560012, India

[c] Laser Materials Development & Devices Division, Raja Ramanna centre for Advanced Technology, Indore 452013, India,

**Abstract**

The Poly(vinylidene fluoride)/$CaCu_3Ti_4O_{12}$ (CCTO) nanocrystal composite films (≈ 85 μm) with relatively high dielectric permittivity ( 90 at 100Hz) were prepared by the solution casting followed by spin coating technique. The structural, the microstructural and the dielectric properties of the composites were studied using X-ray diffraction, Scanning Electron Microscope, and Impedance analyzer respectively. The effective dielectric permittivity ($\varepsilon_{eff}$) of the composite increased with increase in the volume fraction of CCTO at all the frequencies (100 Hz to 1MHz) under investigation. The room temperature dielectric permittivity which is around 90 at 100Hz, has increased to about 290 at 125$^o$C (100Hz). These results may be exploited in the development of high energy density capacitors.

## 1. Introduction

Recently, the high dielectric permittivity composite materials have been considered to be potential candidates for integration into electronic devices. Owing to the continuous development towards the miniaturization of electronics, newer dielectric materials were sought which would enable to achieve high energy density for capacitor applications. Ceramics possessing very high dielectric permittivity are being used as voltage capacitors due to their high breakdown voltages. However, they are brittle, suffer from poor mechanical strength and hence cannot be exposed to high fields. Polymer films such as polyester, polycarbonate, polypropylene, polystyrene and polyethylenesulphide are being used in the fabrication of low leakage capacitors. Though polymers possess relatively low dielectric permittivity, they can withstand high fields, are flexible and easy to process. By combining the advantages of both, one can fabricate new hybrid materials with high dielectric permittivity, and high breakdown voltages to achieve high volume efficiency and energy storage density for applications in capacitors as electric energy storage devices [1-7]. In order to enhance the dielectric permittivity of polymers, ceramic powders such as $Pb(Mg_{1/3}Nb_{2/3})O_3$-$PbTiO_3$(PMNT), $Pb(Zr,Ti)O_3$(PZT), $BaTiO_3$ (BT) [8-13] were used as fillers due to their high dielectric permittivity. Recently, $CaCu_3Ti_4O_{12}$ (CCTO) ceramic which has centrosymmetric *bcc* structure (space group Im3, lattice parameter a ≈ 7.391 $\overset{o}{A}$, and Z=2), has been used as a filler and studied to explore the possibility of obtaining high dielectric permittivity composites for potential capacitor applications [14-21]. It was reported that, the dielectric permittivity as high as 740 at 1 kHz was achieved for a composition of fixed concentration: 50 vol% CCTO and 50 vol% PVDF-TrFE [14]. The dielectric permittivity increases as the CCTO content increases in the polymer and decreases as the frequency increases [15-17]. The reason for increased low frequency dielectric dispersion was attributed to high interfacial polarization triggered by high dielectric loss associated with CCTO [15].

∗ Corresponding author : Tel. +91-80-2293-2914; Fax: +91-80-2360-0683.
E-mail : kbrvarma@mrc.iisc.ernet.in (K.B.R.Varma)

The electrical properties of polymers can be altered / modified by the addition of inorganic nano fillers. Nanoscale particles are more attractive due to their intriguing properties arising from their size associated with large surface area. The insertion of nanoscale fillers may improve the electrical and dielectric properties of the host polymers and the properties can be tailored to a particular performance requirement [22]. But the final properties of a nanocomposite depend on the method of preparation, particle size and the effective dispersion of ceramic particles in the polymer matrix [23-26].

PVDF based composites are being studied in great detail [27-30] because of their better thermal stability, they are tough and can be easily processed by solution cast/injection mould/melt technique. It is also a non-toxic, resistance to heat and chemicals and low water absorption characteristics which make it more suitable for making electronic components. PVDF, a semi-crystalline polymer exists in four different crystalline forms depending on the preparation conditions like solvent, melt temperature, method of casting, stretching of thin films and annealing conditions. The β-phase is the desirable phase owing to its ferroelectric nature. Phase I (β phase) has a planar TTTT (all trans) zigzag chain conformation which has space group Cm2m (orthorhombic, a= 0.858 nm, b= 0.491 nm , c= 0.256 nm) [31-33].

In this work, we report the fabrication and characterization of PVDF/CCTO nano composite system, in which nanocrystallites of CCTO have been dispersed in PVDF solution (dimethyl sulfoxide) followed by spin coating technique for the first time. The composite thus developed has improved dielectric properties which perhaps could be exploited for the development of high energy density capacitors.

## 2. Experimental

*2.1 Characterization*

X-ray powder diffraction (XRD) studies were carried out using an XPERT-PRO Diffractometer (Philips, Netherlands) with Cu K$\alpha_1$ radiation ($\lambda = 0.154056$ nm) in a wide range of 2θ ($5^o \leq 2\theta \leq 85^o$). The microstructure and morphology of the samples were characterized by using scanning electron microscope (FEI Thermal Field Emission SEM Sirion). Transmission Electron Microscopy (TEM) were carried out using FEI-Technai TEM (G-F30, Hillsboro, USA). For the electrical characterization, the films were cut into small pieces of 5X5 mm and gold electrodes with 3mm radii were sputtered at the centre on both sides of each sample. The dielectric studies were carried out using an impedance gain-phase analyzer (HP4194A) as a function of frequency (100Hz–1MHz). The contacts were taken from both sides of the free standing films. The data generated from the instrument was collected through interface between instrument and computer using a software (developed in our laboratory). The measurement accuracy of the instrument is less than 5%. The dielectric permittivity was evaluated using the standard relation $\varepsilon_r = C \times d / \varepsilon_o A$, where $C$ =capacitance, $d$ is the thickness of the sample, $\varepsilon_o$ = 8.854X10$^{-12}$ F/m and A is the effective area of the sample. The dielectric strength measurements were carried out as per the procedure outlined in ASTM D 149.

## 2.2 Preparation of CaCu$_3$Ti$_4$O$_{12}$ nanoparticles.

TiCl$_4$ (titanium tetrachloride, 99.98%) (Merck, Germany), calcium carbonate (BDH; A.R.grade, India), cupric chloride (Fluka, pro analysi grade, India), oxalic acid (s.d.fine-chem Ltd, AR grade, India), NH$_4$Cl and NH$_4$OH(aq) (BDH; A.R.grade, India) ethanol or acetone (Nice chemicals pvt ltd, India; 99.5% pure), dimethylacetamide (DMD) (Merck, Germany) and Poly vinylidene fluoride (PVDF), molecular weight of MW 530,000, supplied by Sigma-Aldrich chemicals pvt ltd, India. CaCu$_3$Ti$_4$O$_{12}$ (CCTO) nanoparticles were synthesized using complex oxalate precursor method [34]. In a typical preparation, titania gel was prepared from the aqueous TiOCl$_2$(0.05M) by adding NH$_4$OH (aq) (at 25$^o$C) till the pH reached ~ 8.0 and NH$_4$Cl was washed off on the filter funnel. This gel was added to 0.4 or 0.8 moles of oxalic acid (2 M solution) (1:1 or 1:2 ratio of Ti: $C_2O_4^{2-}$) which was kept warm (~40$^o$C). To the clear solution obtained, calcium carbonate was added in aliquots and stirred. An aqueous solution containing titanyl oxalic acid together with calcium titanyl oxalate remained clear without any precipitate formation. This solution was cooled to 10$^o$C to which cupric chloride dissolved in acetone along with water (80/20) was added and stirred continuously. The thick precipitate was separated out by further addition of acetone. Subsequently, the precipitate was filtered, washed several times with acetone to make it chloride-free and dried in air. The precursor was isothermally heated around 700°C to get nanocrystals (20-75nm) of phase-pure calcium copper titanate, $CaCu_3Ti_4O_{12}$ as confirmed by X-ray diffraction and TEM studies.

### *2.3 Preparation of PVDF-$CaCu_3Ti_4O_{12}$ nanocrystal composite films*

The composite was prepared by solution casting method. The PVDF polymer was dissolved in dimethyl sulfoxide (DMSO) and an appropriate amount of CCTO nanocrystals was added to the solution, which was thoroughly mixed with the solvent. The suspension was then poured onto a glass plate and then spinned. The free standing composite films of thickness 85μm were obtained and these were annealed at 90 °C for 5 hours which would enable the crystallization of the β-phase of PVDF. The film thus obtained was then heated in a vacuum oven at 80°C for 12 h to remove any remaining traces of the solvent. Composite films with different volume percentages (5% to 30 vol %) of CCTO were prepared.

### 3. Results and Discussion

### *3.1. X-ray diffraction studies*

The X-ray diffraction patterns of PVDF, CCTO and series of PVDF/CCTO composites with different volume percents are shown in Fig.1. The diffraction peaks at 20.7 (200) & 20.8° (110) indicate that the PVDF exists in the β-phase [35]. Fig.1(b) shows the X-ray diffraction pattern obtained for the pure CCTO nano crystalline powder compared well with the ICDD data (01-075-1149) and with the pattern reported earlier [34]. The X-ray diffraction patterns obtained for the PDVF-CCTO (fig.1(c&d)), reflect their composite nature. However, the peak intensity for β-phase of PVDF has decreased as compared to that of CCTO in the composites as the volume percent of CCTO increased in PVDF.

*3.2 Morphology study by SEM/TEM*

Fig. 2 (a&b) shows the bright field TEM image of nano powders of CCTO and the corresponding SAED pattern. Fig 2(a) presents the bright field TEM image of the CCTO nano powders obtained from the oxalate precursor and the size of the crystallite is in the range of 20-75 nm. Fig.2(b) shows the selected area electron diffraction (SAED) pattern with the [012] zone axis. SAED pattern confirms the crystalline nature. The ratio of the reciprocal vectors ($t_2/t_1$) is around 1.229, approaching the calculated value of 1.225 for the bcc lattice.

Fig.3 shows the microstructure of the composite recorded for PVDF+30 vol % CCTO composite and the inset is for the PVDF+ 5 vol % CCTO composite. It is clear that the CCTO crystallites are uniformly distributed in this composite (inset). As the concentration is increased to 30 vol %, the CCTO nano crystallites have the tendency to agglomerate and its size is about 2μm. Though the sizes of the CCTO crystallites remain the same in all the composites, only the size of the agglomerate is different. As revealed by SEM microstructure, the nano particles has the tendency to form clusters, which results in non-uniform distribution of the ceramic powder in the polymer matrix. Therefore, the present work has been restricted to 30 vol % of ceramic powder.

*3.3 Frequency dependence of room temperature dielectric permittivity*

The room temperature dielectric permittivity ($\varepsilon_r^{'}$) and the loss (tan δ) recorded as a function of frequency for PVDF/CCTO nanocomposites are shown in Fig.4 (a & b). The dielectric permittivity (Fig. 4(a)) increases as the ceramic loading increases from 0 to 30 % by volume at all the frequencies under study. It is clearly indicated that the introduction of CCTO nano crystallites in PVDF, increases the dielectric permittivity of the PVDF from 18 to 87 for 30 volume % of CCTO at 100Hz. The dielectric permittivity decreases as the frequency increases from 100Hz to 1MHz. In all the cases, the dielectric permittivity obtained are higher than that of pure PVDF, but much lower than that of pure CCTO [34]. The higher dielectric permittivity obtained in ceramic/polymer composites are attributed to the presence of CCTO nanocrystallites in the PVDF matrix and enhanced polarization from dipole-dipole interaction of closely packed crystallites. The agglomeration formation is attributed to the van der Waals force existing among fine ceramic powders. The dielectric loss (Fig.4(b)) increases with the inclusion of CCTO nanocrystallites in the PVDF matrix. The composite with 30 vol % of CCTO nanocrystallites has the loss value of 0.16 (100Hz). The dielectric loss decreases as the frequency increases. The dielectric loss is considerably higher especially at low frequencies which is mainly attributed to inhomogeneous conduction vis-à-vis interfacial polarization.

### *3.4 Temperature dependence of dielectric properties*

The temperature dependence of dielectric properties of PVDF and PVDF+CCTO-30% composites were studied and illustrated in Figs. 5 and 6 respectively. Fig. 5 shows the frequency dependent dielectric permittivity and loss at different temperatures for pure PVDF. Both the dielectric permittivity and the loss increase with increase in temperature, but decreases as the frequency increases. In the low frequency regime, the dielectric permittivity increased from 24 to 34 when the temperature is increased from $75^{o}C$ to $100^{o}C$ and increased further (to 45) when the temperature is increased to $125^{o}C$. This sudden increase in dielectric permittivity that is observed at 100Hz may be assigned to the space charge/ interfacial polarization effects. The dielectric loss (Fig.5b) has increased from 0.063 to 0.49 as the temperature increased from 30 to $125^{o}C$. Similar observations were reported in the literature for pure PVDF [30].

Fig. 6(a & b) shows the frequency dependence of dielectric permittivity and the dielectric loss for PVDF+30 vol% CCTO composite at different temperatures (30 to $125^{o}C$). The dielectric permittivity increases with increase in temperature but decreases as the frequency increases. The room temperature dielectric permittivity is 87 at 100Hz, which has increased to about 290 at $125^{o}C$ (100Hz). The value at 1kHz is around 65, which has increased to 141 when the temperature is increased from 30 to $125^{o}C$. But the rise in dielectric permittivity with rise in temperature has decreased with increase in the frequency as shown in the Fig.6(a). This behaviour is akin to that of pure PVDF except it has higher dielectric permittivty values as a This behaviour is akin to that of pure PVDF except it has higher dielectric permittivity values as a result of the presence of CCTO nanocrystallites. The most visible change is noticed in the low frequency region (100Hz-10kHz) indicating the strong influence of interfacial polarization mechanism. The inset shows the variation (though it is not that significant as in the previous case) in dielectric permittivity in the 10 kHz to 1MHz frequency range with respect to temperature. The PVDF/CCTO composite exhibits similar dielectric behaviour to that of pure PVDF. The frequency dependent dielectric loss at various temperatures is depicted in the Fig. 6b. The dielectric loss increased from 0.17 to 0.53 as the temperature is increased from 30 to 125°C at 1kHz. At low temperatures, the loss significantly increases subsequent to 1MHz. At higher temperatures the sudden increases in loss may be beyond the frequency range that is covered in the present study.

The effective dielectric permittivity of polymer/filler composite material is dependent not only on the dielectric permittivity of the polymer and the filler, size and shape of the filler and the volume fraction of the filler, but also on the dielectric permittivity of the interphase region, volume of the interphase region and on the type of coupling agents. Hence, it is necessary to predict the dielectric permittivity by combining the theory and the experiment. Various models have been developed for the 0-3 composites [36-39].

The dielectric property of a diphasic dielectric mixture comprising of spherical crystallites with high dielectric permittivity dispersed in a matrix of low dielectric permittivity could be well described by Maxwell's model [36]. According to this model the effective dielectric permittivity of the composite is given by

$$\varepsilon_{eff} = \left( \frac{\delta_p \varepsilon_p (2/3 + \varepsilon_c / 3\varepsilon_p) + \delta_c \varepsilon_c}{\delta_p (2/3 + \varepsilon_c / 3\varepsilon_p) + \delta_c} \right) \qquad (1)$$

where $\varepsilon_c, \varepsilon_p, \delta_c$ and $\delta_p$ and are the dielectric permittivity of CCTO, PVDF, the volume fraction of the dispersoid and the polymer, respectively. Here, the predicted value deviates much from that of the experimental value for all the volume fractions of CCTO under study.

The Maxwell and Furakawa theories were used as the basis for a new theory that was presented by Rayleigh [37]. In this model, the dielectric behaviour of the compsite is given by (Eqn.2)

$$\varepsilon_{eff} = \left( \frac{2\varepsilon_p + \varepsilon_c + 2\delta_c(\varepsilon_c - \varepsilon_p)}{2\varepsilon_p + \varepsilon_c - \delta_c(\varepsilon_c - \varepsilon_p)} \right) \quad (2)$$

Where $\varepsilon_c$ and $\varepsilon_p$ are the dielectric permittivity of the matrix and ceramic particles, respectively, $\varepsilon_{eff}$ is the effective dielectric permittivity and $\delta_c$ is the volume fraction of the ceramic particles. Here, again, it has been observed that the predicted value deviates much from that of the experimental value for all the volume fractions of CCTO under investigation.

The effective medium theory (EMT) model [38] has been established taking into account the morphology of the particles. According to which the effective dielectric permittivity is given by (Eqn.3)

$$\varepsilon_{eff} = \varepsilon_p \left( 1 + \frac{\delta_c(\varepsilon_c - \varepsilon_p)}{\varepsilon_p + n(1-\delta_c)(\varepsilon_c - \varepsilon_p)} \right) \quad (3)$$

where $\delta_c$ is the volume fraction of the ceramic dispersed, $\varepsilon_c$, $\varepsilon_p$ and $n$ are the dielectric permittivity of the particle, polymer and the ceramic morphology fitting factor respectively. The small value of $n$ indicates that the filler particles are of near-spherical shape, while a high value of $n$ indicates largely non-spherical shaped particles. The shape parameter obtained is around 0.49. The effective permittivity of PVDF-CCTO-30 composite computed using the above model for different volume fraction of CCTO is shown in Fig.7a. The experimental data are shown as filled squares.

Fig.7(b) gives the experimental data of permittivity as a function of filler volume fraction. Solid line is fit obtained for the Lichtenecker model (eqn.4) given

$$\log(\varepsilon_c) = V_m . \log(\varepsilon_m) + (0.49) . V_f . \log(\varepsilon_f) \qquad (4)$$

where, $\varepsilon_c, \varepsilon_m$ and $\varepsilon_f$ are the dielectric permittivities of the composite, polymer matrix and the filler, $V_m$ and $V_f$ are the volume fractions of matrix and the filler respectively [19]. It is seen that, the effective permittivity values fitted from these models vary slightly since the shape parameter derived from these models also varies. Thus, it is concluded that the effective dielectric permittivity depends on the shape and size (need to be verified) of the filler particles.

The nano composite films were also studied for dielectric strength as per the procedure outlined in ASTM D 149 and the breakdown tests are carried out in a medium of air. The cylindrical electrodes (both top and bottom) of 6 mm in diameter and the sample was placed between the electrodes and the AC (50 Hz) voltage was continuously increased at a rate of 500 V/s till the sample broke down. Though the measurements were carried out in air, no flashover was noticed. The creepage distance calculated is given in the Table.1. The breakdown voltage, V (kV) of the samples were recorded and the dielectric strength, E (kV/mm) was calculated as E=V/t, where *t* is the thickness of the sample in millimeters. The electrical breakdown data obtained for the PVDF/CCTO nanocomposite films compared with the behaviour of the unfilled PVDF are presented as box and whisker plots (Fig. 8). It is observed that (fig.8), the pure PVDF films has higher electric strength compared to that of the composites. The nanocomposite dielectric strength decreases as the CCTO filler content increases from 5% to 30% by volume in the PVDF. Similar observations were reported for the other PVDF based composite systems [41, 42]. The introduction of fillers into the polymers generally introduces defects in the system causing centers of charge concentration leading to the lower dielectric strength [43]. Hence, the observed trend of decrease in the dielectric strength in the composites is attributed to the CCTO filler and its volume percent in the PVDF. It is also to be noted that the dielectric permittivity has the influence on the dielectric strength. When the breakdown strength is plotted against the dielectric permittivity, an inverse relationship of breakdown voltage to dielectric permittivity is evident [44]. It is also observed that the dielectric permittivity of the PVDF has increased from 18 to 87 as the volume percent of the CCTO increases. Hence, the decreasing trend of electric strength observed from 5% to 30% nanocomposites is attributed to the high dielectric permittivity associated with the CCTO ceramic. Higher the dielectric permittivity associated with higher dielectric loss, act as the channels for charge leakage that lower the dielectric strength in the system.

In order to rationalize the temperature dependence of relaxation processes, electrical modulus approach was adopted. The real ($M'$) and imaginary $M''$ parts of the electrical modulus obtained [45] as a function of temperature at fixed frequency of 5kHz are shown in Fig.9(a and b)

$$M^* = \frac{1}{\varepsilon^*} = \frac{1}{\varepsilon' - j\varepsilon''} = \frac{\varepsilon'}{\varepsilon'^2 + \varepsilon''^2} + j\frac{\varepsilon''}{\varepsilon'^2 + \varepsilon''^2} = M' + j.M'' \quad (5)$$

The $M'$ values decrease as the filler content increases in the PVDF matrix (Fig.9a). The increase of the CCTO content results in lower values of $M'$, implying that the real part of dielectric permittivity increases with ceramic filler. The $M''$ obtained at the same frequency exhibits a relaxation process. The peak maximum value of $M''$ obtained decreases as the filler content increases from 10 to 30 volume percent in PVDF (Fig.9b) which is a characteristic of Maxwell-Wagner-Sillars (MWS) relaxation. Similar observations were made for the CCTO based composite systems where peak maximum value of $M''$ decreases as the filler content increases [19].

Fig.10(a) shows the variation of imaginary part of electrical modulus (M'') at various temperatures as a function of frequency for the PVDF+CCTO-30 composite. Two relaxation processes are clearly observed in the $M''$ curves. The low frequency $M''$ peak shifts towards higher frequency side with rise in temperature, but at high temperature one observes only one relaxation peak. There is a significant change in the relaxation peak height. The intensity of the low frequency relaxation peak is increased and shifts to higher frequency as the temperature increased. This relaxation is attributed to the interfacial or Maxwell-Wagner-Sillars (MWS) polarization which is normally encounted in heterogeneous materials [30]. These relaxation processes were influenced by the interfacial polarization effect which generated electric charge accumulation around the ceramic particles and the shift in the peak position to higher frequencies is attributed to the relaxation phenomena associated with PVDF. Fig.10(b) shows the normalized plots of electric modulus $M''$ versus frequency wherein the frequency is scaled by the peak frequency. A perfect overlapping of all the curves on a single master curve is not found at all the frequencies under study. This indicates that the relaxation process is temperature dependent.

The relaxation time associated with the process was determined from the plot of $M''$ versus frequency. The activation energy involved in the relaxation process is obtained from the temperature-dependent relaxation time ($\tau_{max}$) as

$$\tau_{max} = \tau_o \exp(\frac{E_R}{kT}) \tag{6}$$

where $E_R$ is the activation energy associated with the relaxation process, $\tau_o$ is the pre-exponential factor, $k$ is the Boltzmann constant and $T$ is the absolute temperature. Fig.11 shows a plot between $\ln(\tau)$ and $1000/T(K^{-1})$ along with the theoretical fit (solid line) to the above equation (Eq.(6)). The value that is obtained for $E_R$ is 0.97±0.03 eV is attributed to the relaxation arising from the interfacial polarization.

In Fig. 12, we show the Cole-Cole plot for the PVDF+CCTO-30 composite at various temperatures. In these plots, two distinct semicircles are clearly noticed. The high frequency end semicircle is attributed to the composite nature while the low frequency semicircle is attributed to the interfacial phenomenon occurring between the CCTO particles and the polymer.

## 4.0 *Conclusions.*

High dielectric permittivity Poly(vinylidene fluoride (PVDF) / $CaCu_3Ti_4O_{12}$ (CCTO) nanocrystal composite fils were fabricated. The dielectric permittivity of PVDF increases with increase in CCTO content. The PVDF+CCTO-30% nanocomposite showed higher dielectric permittivity than that of pure PVDF and the other composites under study. The relaxation processes associated with these composites were attributed to the interfacial polarization or MWS effect. Though there is an improvement in the dielectric permittivity, the decrease in dielectric breakdown may limit its use for high voltage applications.

**Acknowledgement**

The management of Central Power Research Institute is acknowledged for the financial support. ( Project No. 5.4.49).

**References.**

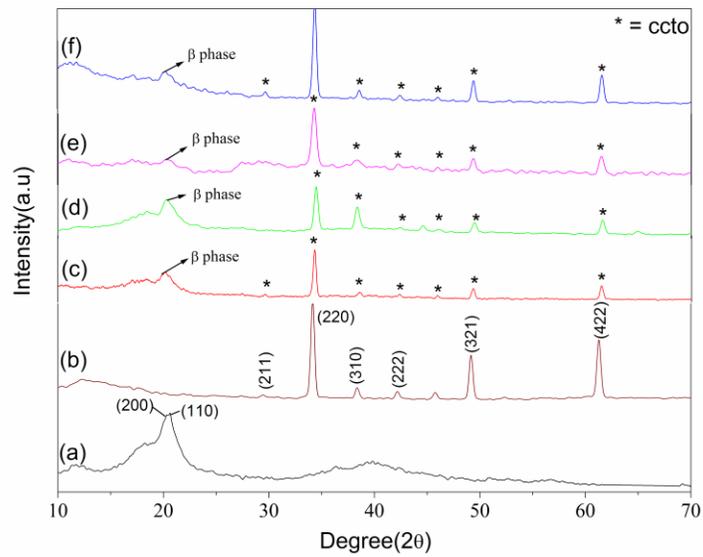

Figure 1. The XRD diffraction patters for : (a) PVDF, (b) CCTO nanocrystalline powder and PVDF-CCTO nanocrysta composites of various concentrations (c) 5 % (d) 10%, (e) 20% and (f) 30vol%

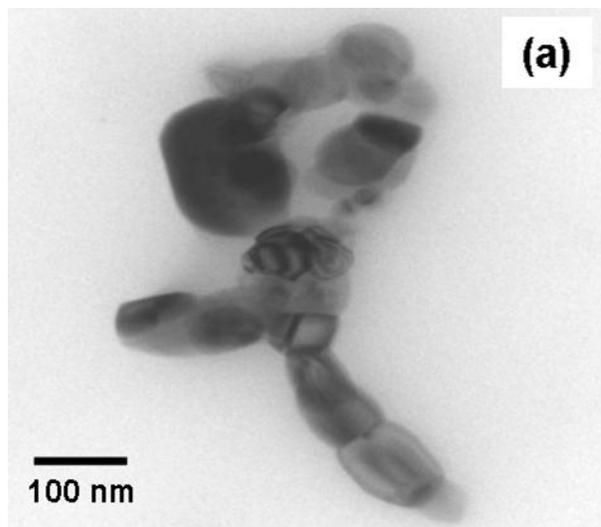

Fig.2 (a). Bright field TEM images of CCTO nanocrystals with dimensions ranging from 20-75 nm

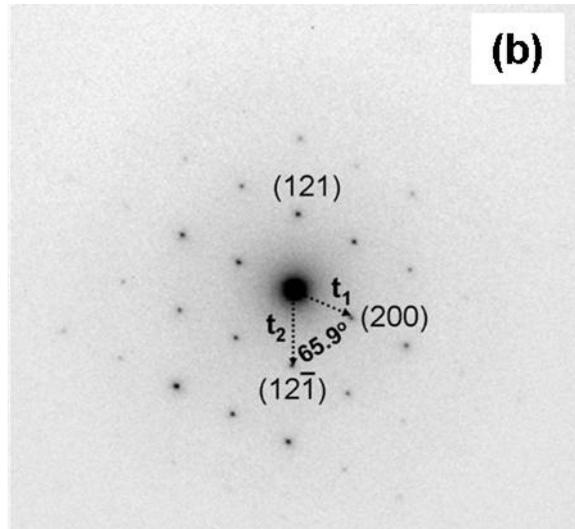

Fig.2 (b) SAED pattern with the zone axis of [012], $t_2/t_1 = 1.229$.

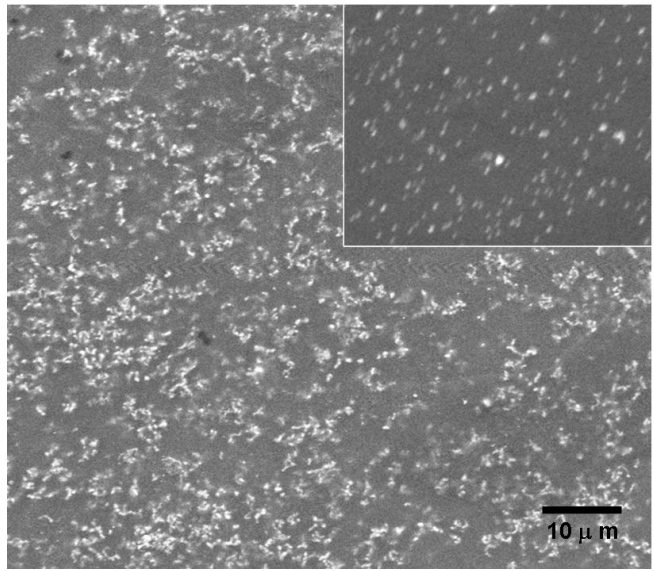

Fig.3. Scanning electron micrographs of PVDF+CCTO-30% nanocrystal composite exhibiting agglomerated CCTO nano crystallite. The inset is for PVDF+CCTO-5% composites with no agglomeration.

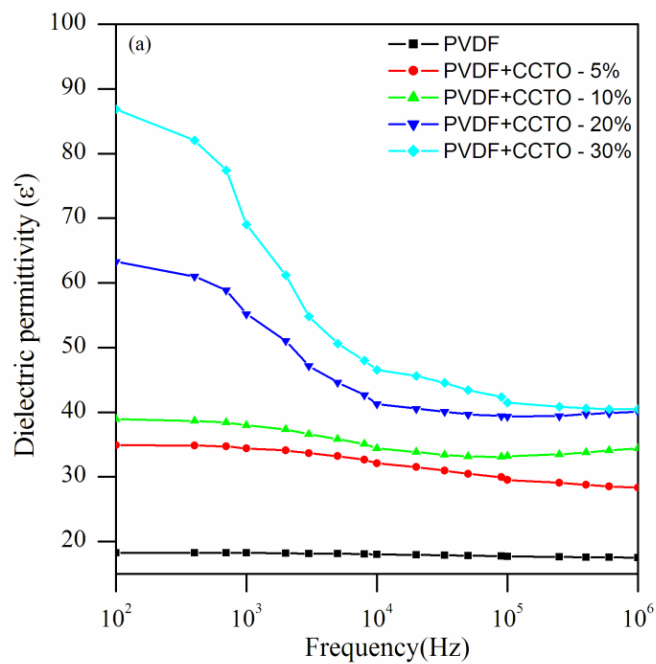

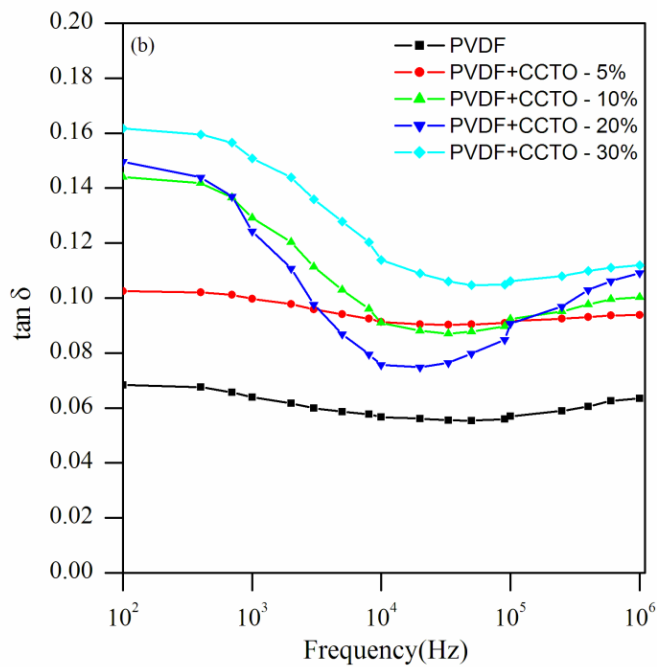

Figure 4. Frequency dependent (a) dielectric permittivity and (b) dielectric loss ( tan δ) of PVDF-CCTO nanocomposite as a function of volume percent of CCTO at 300K.

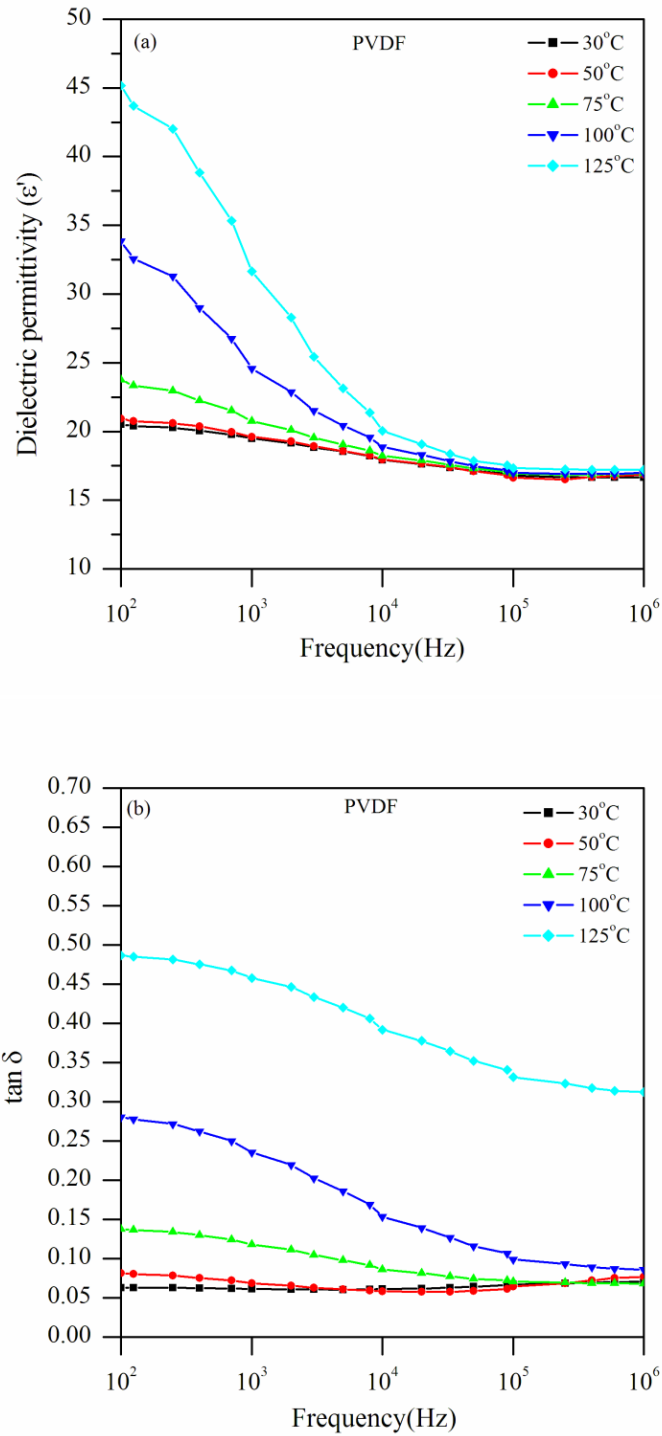

Figure 5. Frequency dependent (a) dielectric permittivity and (b) dielectric loss ( tan δ) at various temperatures for PVDF

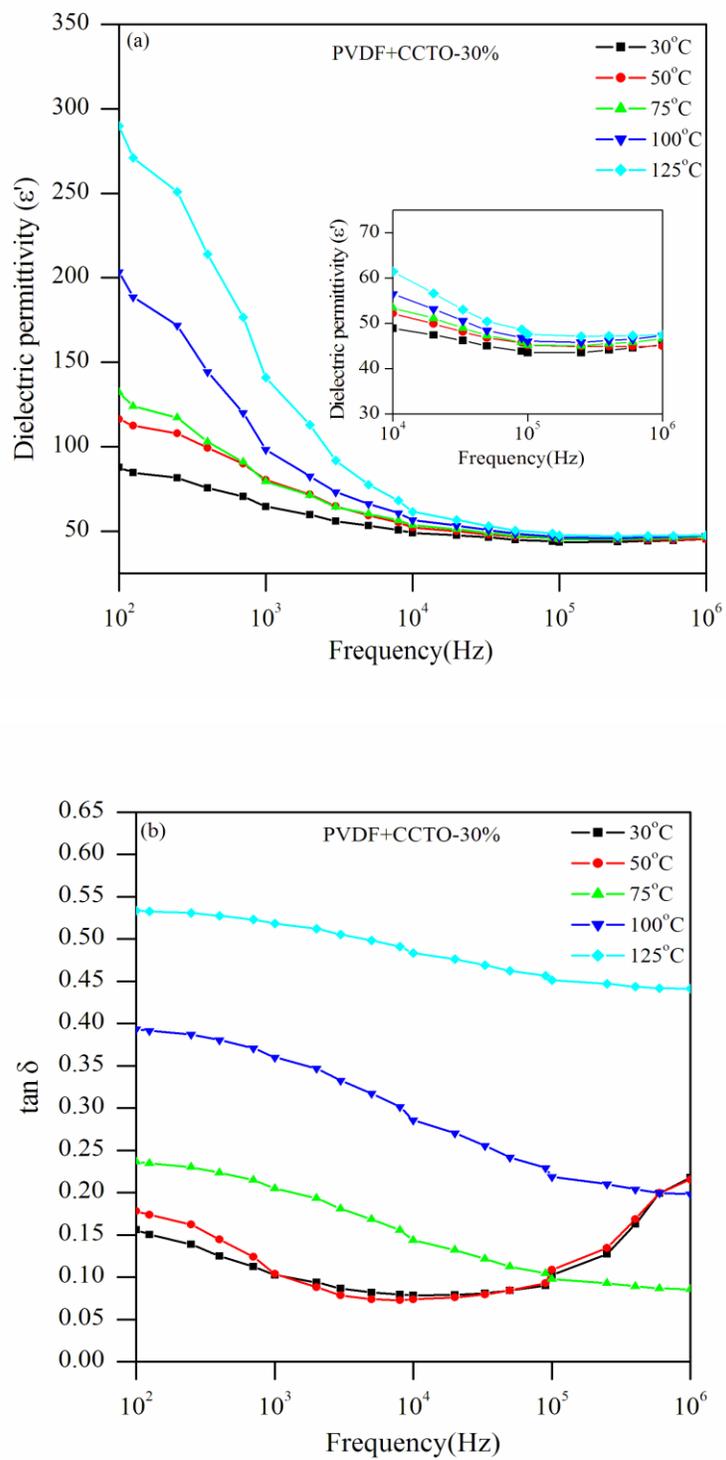

Figure 6. Frequency dependent (a) dielectric permittivity and (b) dielectric loss (tan δ) at various temperatures for PVDF+CCTO-30% nanocomposite.

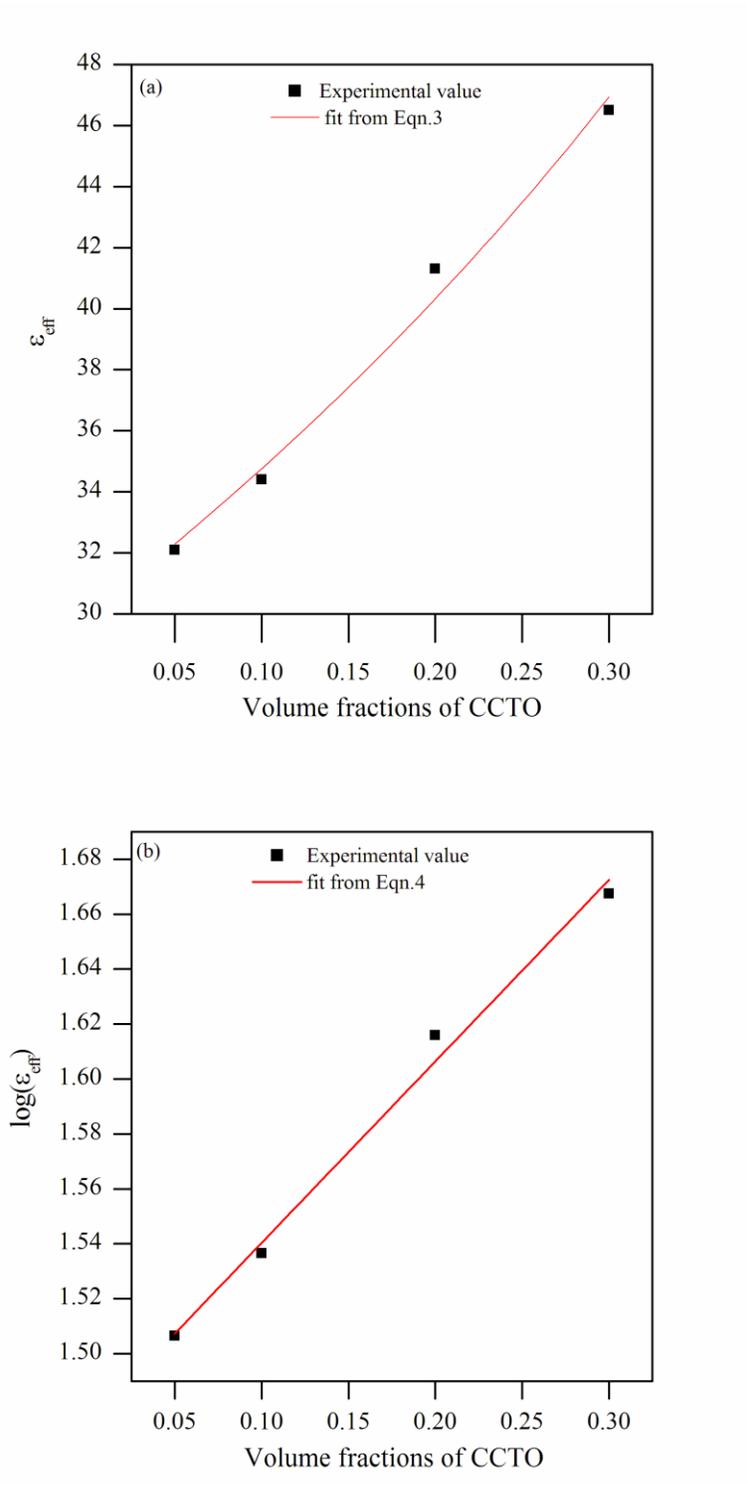

Fig.7. (a) Real permittivity as a function of volume fraction of CCTO. Dots are experimental data and the solid line is fit from the Eqn.(3) (b) fit from the Eqn.(4).

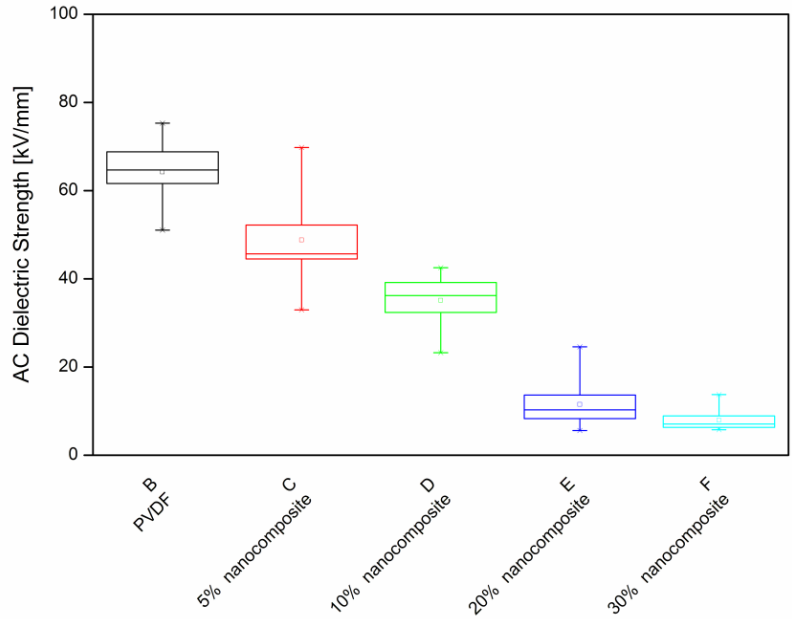

Fig.8. Variation of AC dielectric strength with respect to filler loading in PVDF-CCTO nanocrystal composites.

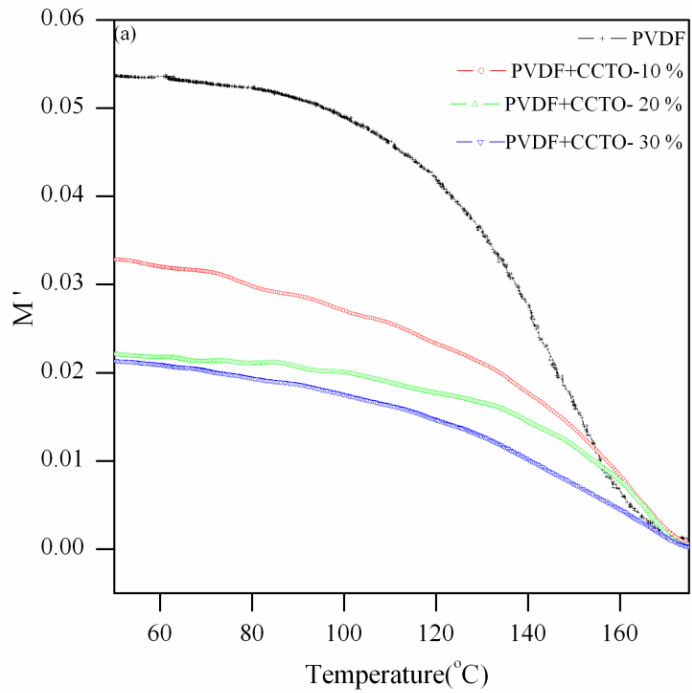

Figure 9. (a) Real part (M´) of electrical modulus vs. temperature for different volume percents of CCTO (at 5kHz).

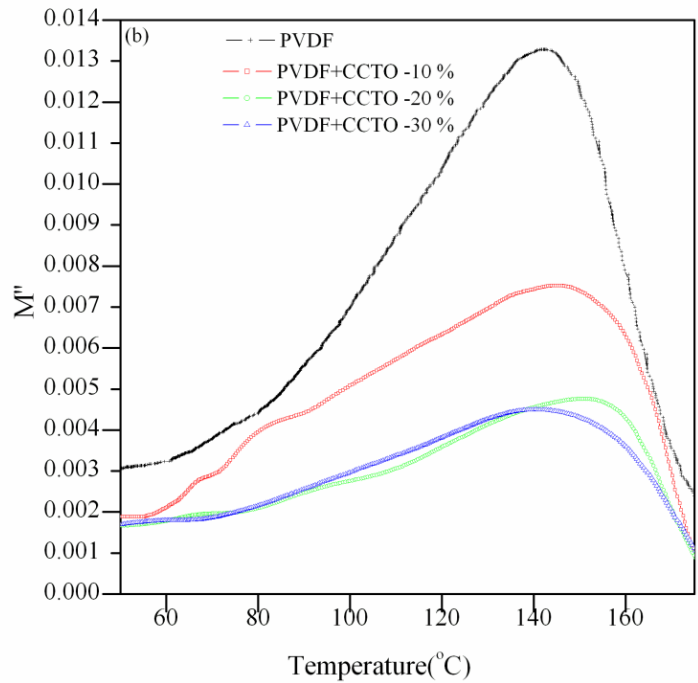

Figure 9. (b) Imaginary part (M´´) of electrical modulus vs. temperature for different volume percents of CCTO (at 5kHz).

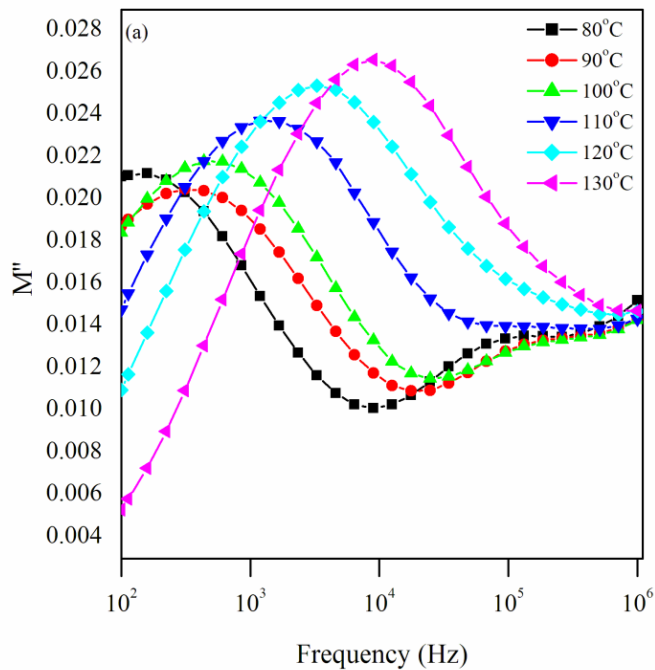

Figure.10 (a). Electric modules spectra for PVDF+CCTO-30% nanocomposite at various temperatures as a function of frequency.

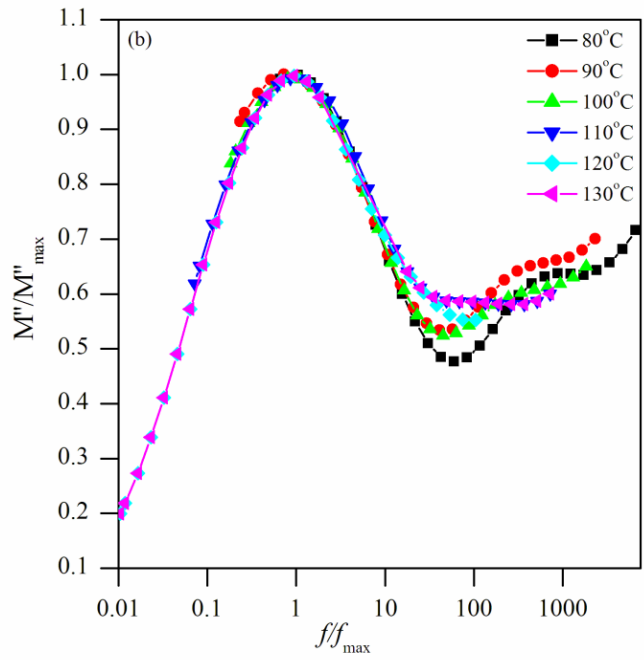

Figure.10(b). Normalized plots of electric modulus against normalized frequency at various temperatures.

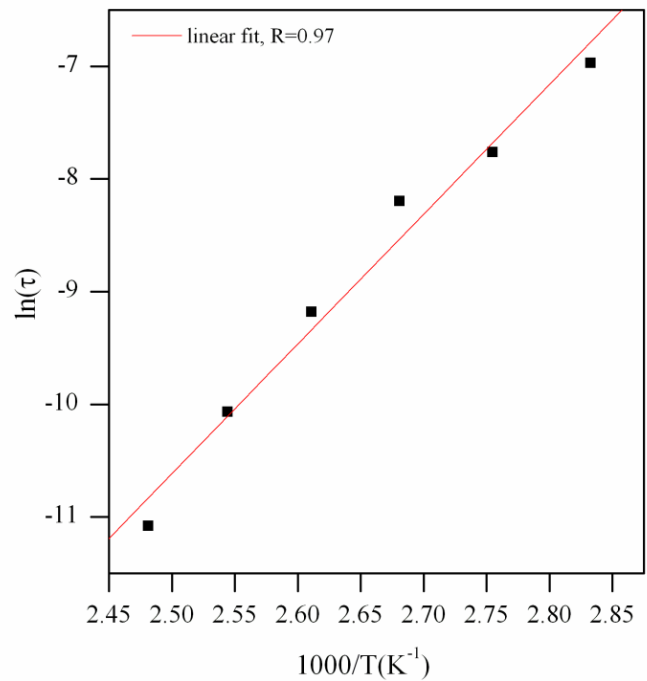

Figure 11. ln ($\tau_{max}$) versus 1000/T (K$^{-1}$)

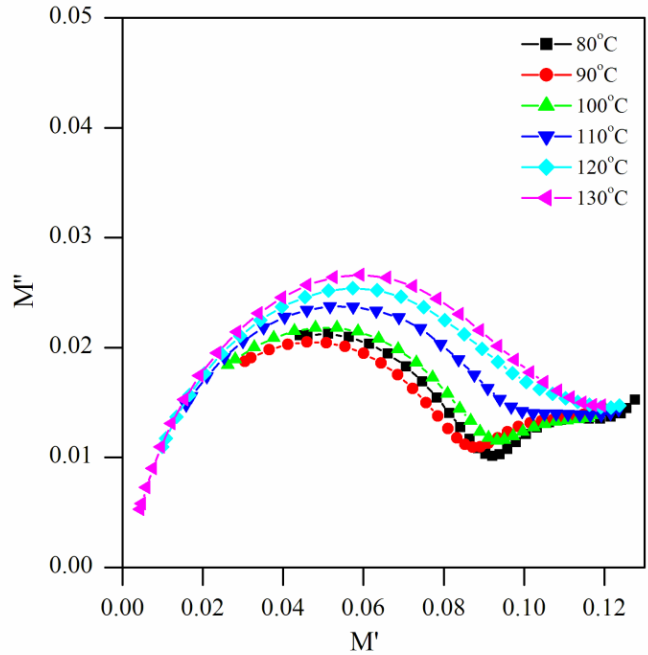

Fig. 12. Cole-Cole plots of the electric modulus, M'' of the PVDF+CCTO-30 composite at various temperatures.

**Table.1. Electric strength and creepage distance for the pure PVDF and PVDF+CCTO nanocrystal composites.**

| Sl.No | Sample details | Electrode diameter | Electric Strength E (kV/mm) | Creepage distance |
|---|---|---|---|---|
| 1 | PVDF film | 6 mm | 64.2 | 16 mm |
| 2 | PVDF+CCTO-5% | 6 mm | 48.8 | 16 mm |
| 3 | PVDF+CCTO-10% | 6 mm | 35.1 | 16 mm |
| 4 | PVDF+CCTO-20% | 6 mm | 11.5 | 16 mm |
| 5 | PVDF+CCTO-30% | 6 mm | 7.9 | 16 mm |